\documentclass[aps,prb]{revtex4}
\usepackage{graphicx}
\usepackage{amsmath,amssymb,bm}

\DeclareMathOperator{\sgn}{sgn}

\DeclareMathOperator{\Tr}{Tr}
\begin{document}
\title{Tunnelling density of states at Coulomb blockade peaks}
%\shorttitle{TDoS at Coulomb blockade peaks}

\author{Nicholas Sedlmayr,  Igor V.~Yurkevich and Igor V.~Lerner}
\affiliation{School of Physics and Astronomy, University of
Birmingham, Birmingham B15 2TT, United Kingdom}

%\shortauthor{N Sedlmayr \etal}

\date{25 July 2006}


\begin{abstract}
We calculate the tunnelling density of states (TDoS) for a quantum dot in
the Coulomb blockade regime, using a functional integral representation
with allowing correctly for the charge quantisation. We show that in
addition to the well-known gap in the TDoS in the Coulomb-blockade
valleys, there is a suppression of the TDoS at the peaks. We show that
such a suppression is necessary in order to get the correct result for the
peak of the differential conductance through an almost close quantum dot.
\end{abstract}

\maketitle

\section{Introduction}
The Coulomb blockade in quantum dots is one of the most thoroughly
investigated and best understood phenomena in modern condensed matter
physics (for reviews see \cite{Alhassid,ABG}). Starting from the earliest
papers on the subject,
\cite{Shekhter,Averin+Likharev:86,GefbenJac,Glazman+MatveevJL:90}, the
attention was focused on the conductance through Coulomb-blockaded dots
and on the corresponding statistics of heights, widths, and positions of
the Coulomb-blockade peaks in conductance. A complimentary approach was to
describe the tunnelling density of states (TDoS) in the dot
\cite{Nazarov:89,KamGef:96,LevShyt:97}. While also leading to the
description of conductance and its distribution it was of considerable
interest by itself. The reason is that a singularity in the TDoS in the
zero-dimensional dot potentially bridges the zero-bias anomaly in the TDoS
of interacting electrons in the metallic regime \cite{AA:79} with the
Coulomb gap in the insulating regime \cite{Shklovskii+Efros:75}.

Qualitatively, behaviour of the TDoS found in
\cite{Nazarov:89,KamGef:96,LevShyt:97} has a clear origin. The Coulomb
blockade makes the energetic cost of putting an extra electron in the dot
of the order of the charging energy, $E_{\text{c}}=e^2/C$ ($C$ is an
effective capacitance). Consequently, it leads to a low-energy gap in the
TDoS stretching up to $E_{\text{c}}$ at low temperatures, $T\ll
E_{\text{c}}$. By tuning the gate voltage applied to the dot  one
periodically reaches degeneracy points where the energy of having $N$ and
$N+1$ electrons in the dot is the same. At these points, i.e.\ at the
peaks in the Coulomb blockade regime, tunnelling is no longer suppressed
and one expects the gap in the TDoS to  close.  In this Letter we show
that, although the TDoS at the peaks remains finite, its zero-energy value
is of half of its value at $\varepsilon \gtrsim E_{\text{c}}$. We then
demonstrate that such a suppression is necessary to restore the correct
results \cite{Shekhter,Averin+Likharev:86,GefbenJac,Glazman+MatveevJL:90}
for the peak conductance.

We will use technique similar to that introduced by Kamenev and
Gefen \cite{KamGef:96} in their original calculation of the TDoS
in the quantum dot, representing the electron Green's functions as
functional integrals, albeit we use the Keldysh representation in
the form developed in \cite{KamAn:99} rather than the Matsubara
one as in \cite{KamGef:96}. The main difference in our approach is
that in the saddle-point approximation, we take into account all
the electron winding numbers. This is necessary to get the correct
expression for the TDoS at the Coulomb blockade peaks and also is
the only consistent way (in the discussed context) to get the
electron number  quantization both at the peaks and in the
valleys. Naturally, this could be obtained either in Keldysh or in
Matsubara formalism. The necessity of including all the winding
numbers has already been stressed in \cite{Efetov+Tschersich:02}
for the  problem of a granular medium comprised of an array of
quantum dots in the Coulomb blockade regime. However, the
effective averaging over chemical potentials of the dots
inevitable for the granular system, gives in this case only the
value of the TDoS in the valleys.

\begin{figure}
\begin{center}
  \includegraphics[width=.5\textwidth]{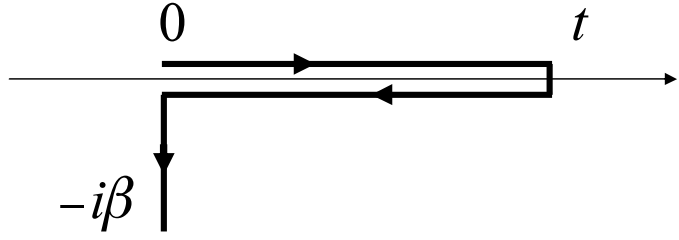}\\
  \caption{The `interaction' Keldysh contour \cite{RS:86}.}\label{1}
\end{center}
\end{figure}

We consider the problem in the ergodic (zero-dimensional) limit. It is
straightforward in our formalism to include a correction from spatially
inhomogeneous modes; however, this correction is small as $1/g$ (where
$g\sim 1/\tau_{\text{dw}}\delta\gg1$ is the dimensionless conductance of
the dot and $\tau_{\text{dw}}$ is the electron's mean dwelling time in the
dot) and will not differ from that found previously \cite{KamGef:96}. Thus
we start with the standard `universal' Hamiltonian for a zero-dimensional
system \cite{KamGef:96,Kurland,ABG}, keeping there only the charging term:
\begin{align}\label{H}
\hat H   =\hat H_0  +\frac{E_{\text{c}}}{2}\left({\hat N-N_g}\right)^2.
\end{align}
$\hat{H}_0$ is the Hamiltonian of free electrons in a random
potential $V$, $\hat{N}$ is the electron number operator, $eN_{ {g}}$
is
 the neutralising background charge (governed by the gate voltage
for the
 standard quantum dot). It is convenient to represent the
appropriate Green's function in terms of the functional integral in
the Keldysh technique
\cite{KamAn:99},
\begin{align}\label{G}
iG({\mathbf r},t;{\mathbf r}',t')={{{Z}}}^{-1}\int
 {\mathcal{D}}\bar{\psi} {\mathcal{D}}\psi\,
 \psi({\mathbf r},t)\bar\psi({\mathbf r}',t')\, {\mathrm e}^{iS[\psi]}\,,
\end{align}
where $Z=\int
 {\mathcal{D}}\bar{\psi} {\mathcal{D}}\psi\,
   {\mathrm e}^{iS[\psi]}$
 and the action $S[\psi] =S_0[\psi]+S_{\text{c}}[\psi] $ is given by
\begin{equation}\label{S}
    \begin{aligned}
S_0[\psi]&=\int_K{\mathrm d} t\int{\rm d}{\mathbf r}\,\bar{\psi
}({\mathbf r},t)\left[i\partial_t -\hat\xi\right]\psi({\mathbf
r},t)\,,
 \qquad \hat{\xi}\equiv \frac{\hat{p}^2}{2m}+V-\widetilde{\mu} \\
S_{{\text{c}}}[\psi]&=-\frac{E_{\text{c}}}{2}\int_K{\mathrm d} t\,
N^2(t), \qquad N(t)=\int{\mathrm d} {\mathbf r}\,
\bar\psi({\mathbf r},t)\psi({\mathbf r},t)\,.
\end{aligned}
\end{equation}
The time integrals are taken along the standard `interaction'
contour\cite{Keld,RS:86}, Fig.\ref{1},  and
$\widetilde{\mu}=\mu+E_{\text{c}}  {N_g}$ on the thermodynamic
(imaginary) part of the contour and
$\widetilde{\mu}=E_{\text{c}}N_g$ on its dynamical (real) part.
This corresponds to a Green's function defined as an average with
the grand canonical Gibbs density matrix; its time evolution is
described in the Heisenberg representation with a Hamiltonian,
Eq.~(\ref{H}), that does not contain $\mu$. In the
zero-dimensional regime considered here it is convenient to expand
Grassmann fields $\psi({\mathbf r},t)$ and $\bar{\psi}({\mathbf
r},t) $ in terms of free electron eigenfunctions, introducing
fields $c(t)$ and $\bar{c}(t)$ which depend on time only:
\begin{align*}%\label{decomp}
\psi({\mathbf r},t)=\sum_n\psi_n({\mathbf r})\,c_n(t), \quad
\hat\xi\psi_n({\mathbf r})=\xi_n\,\psi_n({\mathbf r}), \quad
\xi_n=\varepsilon_n-\widetilde{\mu}.
\end{align*}
The Green's function becomes
\begin{align}\label{GF}
G({\mathbf r},t;{\mathbf r}',t')=\sum_n\psi_n({\mathbf
r})\bar\psi_n({\mathbf r}')\,G_n(t,t')
\end{align}
In the zero-dimensional system under consideration positions
$\mathbf{r}$ and $\mathbf{r}'$ in Eq.~(\ref{GF}) are indistinguishable so
that observable quantities can be found from
\begin{align}\label{Gr}
    G(t,t')\equiv\int\! \mathrm{d} ^d r\,G({\mathbf
r},t;{\mathbf r},t')=\sum_{n}^{ }G_n(t,t')\,.
\end{align}
Decoupling the charging term $S_{\text{c}}$ in Eq.~(\ref{S}) with
the help of the standard Hubbard-Stratonovich transformation leads
to (i) replacing $S_{\text{c}}$ by the bosonic action
\begin{align}\label{phi}
S[\phi]=-\frac{1}{2E_{\text{c}}}\int_{K}{\mathrm d}
t\,\phi^2(t)\,,
\end{align}
and (ii) substituting $i\partial_t-\phi$ for $i\partial_t$ in the
 action $S_0$. To calculate the functional integral over
the fermionic fields $c_n(t)$ and $\bar{c}_n(t)$, and thus $G(t,
t')$, we notice that for any function $\varphi(t)$
\begin{align*}
Z^{-1}_0\int{\mathcal D}\bar{c}_n\,{\mathcal D}c_n\,{\mathrm
e}^{i\int_K{\mathrm d} t\,\bar c_n\left[i\partial_t
+\varphi(t)\right]c_n}&=1+{\mathrm e}^{i\int_{K}{\mathrm d}
t\,\varphi(t)},
\\
Z^{-1}_0\int{\mathcal D}\bar{c}_n\,{\mathcal D}c_n\,c_n(t)\bar c_n(t')\,
{\mathrm e}^{i\int_K{\mathrm d} t\,\bar c_n\left[i\partial_t
+\varphi(t)\right]c_n}&={\sgn}(t,t')\,{\mathrm e}^{i\int _{t'}^{t}{\mathrm
d} \tau\,\varphi(\tau)}\,.
\end{align*}
The time ordering here is along the contour in Fig.~\ref1 so that
the integral in the second line above is also taken along the
contour  and ${\sgn}(t,t')$ equals $1$ (or $-1$) when
 $t$ precedes (or goes after) $t'$ on the
contour. The bosonic field is not included in the normalization
factor $Z_0$ in the above expressions.

Note the analogy with the calculations in the Matsubara technique
in Ref. \cite{KamGef:96}: the  bosonic field can be gauged out by
a shift in the fermionic field but for the initial conditions in
the imaginary time integral. These do not allow one to get rid in
this way of the zero-frequency field component. In our
calculations $\phi_0$ is the precise analogue of this component.
Hence
\begin{align}\label{gn}
 G_n(t,t')&=\frac{-i\sgn(t,t')} {\mathcal{Z}}  \int{\mathcal D}\phi\, {\mathrm
e}^{iS[\phi]}\,\Xi_n(\phi_0)\,{\mathrm
e}^{\int_{t'}^{t}{\mathrm{d}}\tau[-i\xi_n+\phi(\tau)]}\,, &
{{\mathcal{Z}}}&=\int{\mathcal D}\phi\, {\mathrm e}^{iS[\phi]}\,\Xi(\phi_0) \,.
\end{align}
Here we have defined the grand canonical partition function, $\Xi(\phi_0)$, and
the grand canonical partition function with the $n^{\text{th}}$ level excluded,
$\Xi_n(\phi_0)$, with energy levels in both shifted by the charging effects:
\begin{align}\label{Xi}
\Xi(\phi_0)&\equiv\prod_m\left(1+{\mathrm e}^{-\beta\xi_m+\phi_0}\right)\,,&
\Xi_n(\phi_0)& \equiv \Xi(\phi_0) \left(1+{\mathrm
e}^{-\beta\xi_n+\phi_0}\right)^{-1}\,.
\end{align}
It is convenient to expand $\Xi(\phi_0)$ and $\Xi_n(\phi_0)$ in
Eq.~(\ref{Xi}) in terms of the canonical partition functions:
\begin{equation}\label{Z}
\begin{aligned}
\Xi(\phi_0)&=\sum_{N=0}^{\infty}\,Z_N\,{\mathrm
e}^{(\beta\widetilde{\mu}+\phi_0)N}\,, &Z_N&=\oint\frac{\rm
d\theta}{2\pi}\,{\mathrm e}^{-iN\theta}\,\prod_m\left(1+{\mathrm
e}^{-\beta{\varepsilon}_m+i\theta}\right)\,,\\
\Xi_n(\phi_0)&=\sum_{N=0}^{\infty}\,Z_N(\varepsilon _n)\,{\mathrm
e}^{(\beta\widetilde{\mu}+\phi_0)N}\,, &Z_N(\varepsilon _n)&=\oint\frac{\rm
d\theta}{2\pi}\,{\mathrm e}^{-iN\theta}\,\prod_{m\ne n}\left(1+{\mathrm
e}^{-\beta{\varepsilon}_m+i\theta}\right)\,.
\end{aligned}
\end{equation}
The meaning of $Z_N({\varepsilon}_n)$ can be seen from its formal definition:
\begin{align}\label{Z/Z}
    \frac{Z_N(\varepsilon_n)}{Z_N} = \frac{\Tr_N\Big( {c_nc_n^\dagger \mathrm{e}}
    ^{-\beta\hat{H}_0} \Big)}{\Tr_N\left( {  \mathrm{e}}
    ^{-\beta\hat{H}_0} \right) }=1-F_N(\varepsilon_n)\,,
\end{align}
where $F_N(\varepsilon_n)$ is the distribution function in the system of
$N$ noninteracting electrons, as the charging energy in the Hamiltonian
(\ref{H}) is just a constant when the number of electrons $N$ is fixed.
Note that in order to get a consistent description of the charge
quantization in the quantum dot, it is crucial to have all winding numbers
$m$ in the expansion (\ref{Z}).

On substituting this expansion and a similar one for $\Xi_n$ into
Eq.~(\ref{gn}), and performing the Gaussian integration over the
fields $\phi$, we find  all the Keldysh components of the Green's
function. Thus,  the Fourier transform of $G^>(t,t')$ (where $t$
is on the lower, and $t'$ is on the upper
 part of the Keldysh
contour, Fig.\ref{1}) is given by:
\begin{align}\label{Gn>}
    G^>(\varepsilon)&= -\frac{2\pi i}{\cal
Z}\sum_{n}^{}\sum_{N=0}^{\infty}\,e^{-\beta E_N}\,Z_N(\varepsilon- \Omega_N)\,
\delta(\varepsilon-\varepsilon_n-\Omega_N)\,,
\end{align}
where
\begin{align*}
    {\cal Z}&= \sum_{N=0}^{\infty}\,\mathrm{e}^{-\beta E_N}\,Z_N\,,&
E_N&\equiv\frac{E_{\text{c}}}{2}\,\left( {N-N_g} \right)^2 - \mu\,N\,,
&\Omega_N&\equiv E_{\text{c}}\left(N+\tfrac{1}{2}-{N_g}\right).
\end{align*}
Now we can  effectively average over disorder simply by substituting the mean
TDoS of noninteracting electrons,
$
    \nu_0$, for $\sum_{n}^{ }\delta(\varepsilon-\varepsilon_n-\Omega_N)\,,
$ with the assumption that the TDoS is smooth in any realisation of
disorder which is valid when the mean (or typical) level spacing $\delta$
is much smaller than $T$. Next we use Eq.~(\ref{Z/Z}) to obtain
\begin{align}\label{G>}
    G^>(\varepsilon)&= -\frac{2\pi i\nu_0}{\cal Z}\sum_{N}\,\mathrm{e}^{-\beta
E_N} \left[1-F_N(\varepsilon-\Omega_N)\right]\approx -\frac{2\pi i\nu_0}{\cal
Z}\sum_{N}\,\mathrm{e}^{-\beta E_N}\left[1-f(\varepsilon-\Omega_N)\right],
\end{align}
where we have used that $F_N(\varepsilon-\Omega_N)$ for $N\gg1$ is
approximately the same as the (grand canonical) Fermi-Dirac distribution
function $f(\varepsilon-\Omega_N)$ with the `chemical potential' $\mu_0$ of
order $N\delta$ (which is small as compared to $\Omega_N$ and thus discarded).
Note that this obvious relation between canonical and grand canonical
expressions can be easily confirmed directly by using the definition
(\ref{Z/Z}) and calculating $Z_N$ and $Z_N(\varepsilon _n)$, Eq.~(\ref{Z}), in
the saddle point approximation.

\begin{figure}
\begin{center}
$\begin{array}{c@{\hspace{0.2in}}c@{\hspace{0.2in}}c}
  \includegraphics[width=0.31\textwidth]{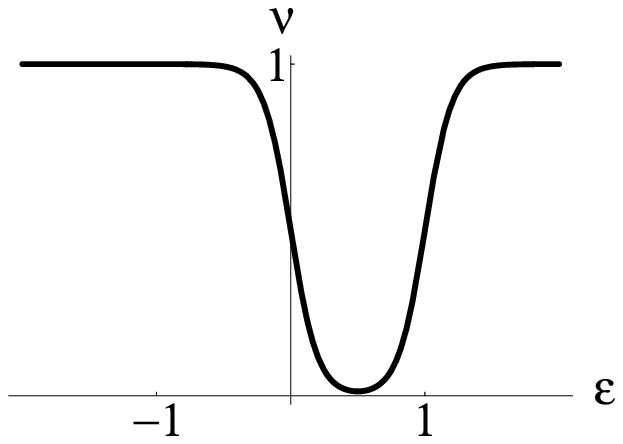}&
  \includegraphics[width=0.31\textwidth]{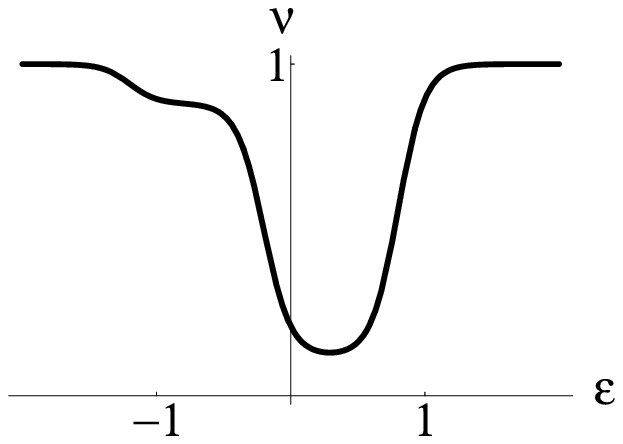}&
  \includegraphics[width=0.31\textwidth]{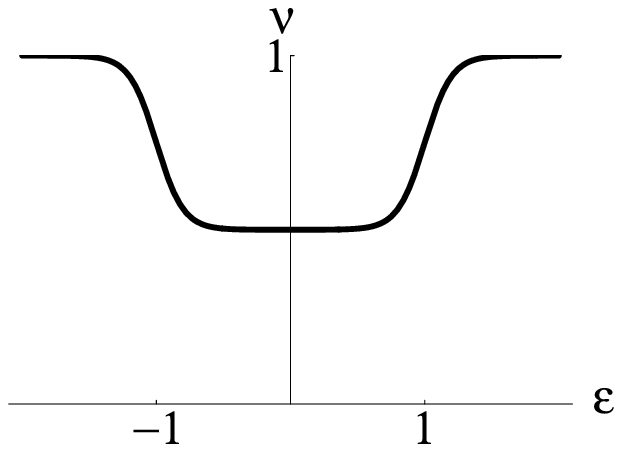}
\\ \textrm{(a)}&\textrm{(b)}&\textrm{(c)}
\end{array}$
\caption{The dependence of the TDoS (in the units of $\nu_0$) on
the energy (measured in $E_{\text{c}}$):  (a) in the valley, (b)
through an intermediate region, and  (c) at the peak.}\label{2}
\end{center}
\end{figure}

The TDoS can be found from the standard formula
\begin{align}\label{nu}
    \nu(\varepsilon )=\frac{i}{2\pi} [G ^R(\varepsilon )-G ^A(\varepsilon )]
    =\frac{i}{2\pi} [G ^>(\varepsilon )-G ^<(\varepsilon )]\,.
\end{align}
Although $G^<$ and $G^>$ are exactly related (in the equilibrium case)
by
 $G^<_n(\varepsilon )=-\mathrm{e}^{\beta(\varepsilon
-\mu)}G^>_n(\varepsilon )$, this is not   convenient to use for
approximations in the strong Coulomb blockade
 regime
($E_{\text{c}}\beta\gg1$) as both functions in the product are very
sharp. Instead, we note that there exists an exact relation,
$Z_N=Z_N(\varepsilon_n)+e^{-\beta\varepsilon_n}\,Z_{N-1}(\varepsilon_n)$
that
 follows from the definition (\ref{Z}), which allows us to
express $G^<$ after
 straightforward transformations as
\begin{align}\label{G<}
    G^<(\varepsilon)&= \frac{2\pi i\nu_0}{\cal Z}\,\sum_{N}\,e^{-\beta
E_{N}}\,f(\varepsilon-\Omega_{N-1})\,.
\end{align}
From these expressions for the Green's functions we find the TDoS to be:
\begin{align}
\frac{\nu(\varepsilon)}{\nu_0}=\frac{1}{\cal Z}\sum_N\,e^{-\beta
E_N}\left[f\left(\varepsilon-E_c(N-\tfrac12-N_g)\right)+
1-f\left(\varepsilon-E_c(N+\tfrac12-N_g)
\right)\right].
\end{align}
Keeping only the leading terms in the sum over $N$, we get
\begin{align}\label{16}
\frac{\nu(\varepsilon)}{\nu_0}=\frac{U(\varepsilon-\Omega_N)
+e^{-\beta(\Omega_N-\mu)}\,
U(\varepsilon-\Omega_{N+1})}{1+e^{-\beta(\Omega_N-\mu)}} ,
\end{align}
where we have defined $
U(\varepsilon-\Omega_N)=f(\varepsilon-\Omega_{N-1})+1-f(\varepsilon-\Omega_N)
$. All other terms are exponentially suppressed so that we just
have the terms for $N$ closest to $N_g+  \tfrac12$. Away from the
degeneracy point,  one of the terms in eq.~(\ref{16}) is also
exponentially suppressed so that this expression is contributed by
one term only. This corresponds to the valley in the Coulomb
blockade and gives the TDoS with a gap, as illustrated in fig.
\ref{2}(a). In approaching the degeneracy point, fig. \ref{2}(b),
the gap is smeared by the contribution from the other term in
eq.~(\ref{16}); finally, at the degeneracy point corresponding to
the peak of the Coulomb blockade, the TDoS remains finite at all
energies but shows a `half-gap' at ${\varepsilon}<E_{{\text{c}}}$,
fig. \ref{2}(c).

Now we show that the half-gap in the TDoS is necessary to restore
a correct expression for the current through an almost closed
quantum dot.  We consider such a dot   connected via weak
tunnelling contacts to leads (reservoirs) with fixed chemical
potentials. Introducing in the usual way \cite{RS:86} the
retarded, advanced and Keldysh components of the Green's function,
one can rewrite  the standard expression \cite{JWM:94} for the
current through the $\alpha^{{\text{th}}}$ lead (with
$\Gamma_\alpha=2\pi \nu_\alpha |\gamma_\alpha|^2$ being the
tunnelling rate, and $\gamma_\alpha$ tunnelling coefficients) as
follows:
\begin{align}\label{current}
\mathcal{I}_\alpha=e\Gamma_\alpha\int_{-\infty}^{+\infty}\frac{{\mathrm
d}\varepsilon}{4\pi i}\,\Tr \left\{\hat
G^K(\varepsilon)-\big[{1-2f_\alpha({\varepsilon}) }\big]
\big[G^R(\varepsilon)-G^A(\varepsilon)\big]\right\}\,.
\end{align}
Restricting considerations to the case of two leads ($\alpha=1,2$)
and using the current conservation,
$\mathcal{I}_1+\mathcal{I}_2=0$, we exclude the Keldysh compponent
$G^K$ from eq.~(\ref{current}) to find
\begin{align}
\mathcal{I}_\alpha
=\frac{e\Gamma_1\Gamma_2}{\Gamma}\int_{-\infty}^{+\infty}\frac{{\rm
d}\varepsilon}{2\pi
i}\big[f_{2}(\varepsilon)-f_{1}(\varepsilon)\big]
\big[G^R(\varepsilon)-G^A(\varepsilon)\big]
\end{align}
In the linear response regime, when the difference between the
chemical potentials in the leads $\mu_2-\mu_1=eV$ , we obtain the
following expression for the differential conductance,
\begin{align}\label{dIdV}
\frac{{\mathrm d}\mathcal{I}}{{\mathrm d} {V}}
=e^2\,\nu_0\frac{\Gamma_1\Gamma_2}{\Gamma}\int_{-\infty}^{+\infty}{\rm
d}\varepsilon\left[-\frac{\partial
f(\varepsilon-\mu)}{\partial\varepsilon
}\right]\,\frac{\nu(\varepsilon)}{\nu_0}
\end{align}
Here $\nu({\varepsilon})$ is the TDoS calculated for   the dot in
a thermodynamical contact with the leads at the same chemical
potential $\mu$.  Substituting the expressions for the TDoS
obtained above and keeping only the two leading order terms $N$
and $N+1$ in the TDoS, i.e.\ using eq.~(\ref{16}), we arrive at
\begin{align*}
\int_{-\infty}^{+\infty}{\rm d}\varepsilon\left[-\frac{\partial
f(\varepsilon-\mu)}{\partial\varepsilon
}\right]\,\frac{\nu(\varepsilon)}{\nu_0}=\frac{\Omega_N/2T}{\sinh\Omega_N/2T}\,,
\end{align*}
provided that $|\Omega_N|=|E_a(N+\tfrac12)- {N_g}|\ll E_a$.
Substituting this into eq.~(\ref{dIdV}) reproduces the standard
result \cite{Alhassid,ABG} for the differential conductance:
\begin{align}
\frac{{\mathrm d}\mathcal{I}}{{\mathrm d} {V}}
=e^2\,\nu_0\frac{\Gamma_1\Gamma_2}{\Gamma}\,
\frac{\Omega_N/2T}{\sinh\Omega_N/2T}.
\end{align}
We note that for  reproducing the correct coefficient at the peak
in the differential conductance  it is necessary to have the
half-gap structure in the TDoS described above.

In conclusion, we have calculated the Green's functions describing
an isolated quantum dot in the Coulomb blockade regime. This
allows us to find a complete description of the TDoS both in the
Coulomb valleys and peaks. At the peaks we find a new feature in
the TDoS: the zero-energy suppression of the TDoS  by the factor
of 2 in comparison to its value at high energies. By considering
the quantum dot weakly coupled to two leads we can find the linear
response conductance for elastic tunnelling. We have also
demonstrated that the correct description of the TDoS at the peak
is necessary to obtain the standard result for the differential
conductance. Finally, we believe that the techniques developed
here could be extended for considerations of the spin blockade
effects in the quantum dot \cite{KisGef:06}.

\acknowledgements{This work was supported by the EPSRC grant
GR/R95432.}

%\bibliography{my} \end{document}

%\bibliography{my}

\begin{thebibliography}{10}

\bibitem{Alhassid}
\textsc{Alhassid Y.}, \emph{Rev. Mod. Phys.} \textbf{72} (2000)
895–.

\bibitem{ABG}
\textsc{Aleiner I.~L.}, \textsc{Brouwer P.~W.} and \textsc{Glazman
L.~I.},
  \emph{Phys. Rep.} \textbf{358} (2002) 309.

\bibitem{Shekhter}
\textsc{Kulik I.~O.} and \textsc{Shekhter R.~I.},  \emph{Zh.\
Eksp.\ Teor.\
  Fiz.} \textbf{68} (1975) 623.

\bibitem{Averin+Likharev:86}
\textsc{Averin D.~V.} and \textsc{Likharev K.~K.}, \emph{J. Low
Temp. Phys.}
  \textbf{62} (1986) 345.

\bibitem{GefbenJac}
\textsc{{Mullen} K.}, \textsc{{Gefen} Y.} and \textsc{{Ben-Jacob}
E.},
  \emph{Physica B} \textbf{152} (1988) 172.

\bibitem{Glazman+MatveevJL:90}
\textsc{{Glazman} L.~I.} and \textsc{{Matveev} K.~A.} \emph{JETP
Lett.}
  \textbf{51} (1990) 484.

\bibitem{Nazarov:89}
\textsc{Nazarov Y.~V.},  \emph{Zh.\ Eksp.\ Teor.\ Fiz.}
\textbf{96} (1989) 975.

\bibitem{KamGef:96}
\textsc{Kamenev A.} and \textsc{Gefen Y.}, \emph{Phys. Rev. B}
\textbf{54}
  (1996) 5428.

\bibitem{LevShyt:97}
\textsc{Levitov L.~S.} and \textsc{Shytov A.~V.} \emph{JETP Lett.}
\textbf{66}
  (1997) 214.

\bibitem{AA:79}
\textsc{Altshuler B.~L.} and \textsc{Aronov A.~G.}, \emph{Sol.
State Commun.}
  \textbf{30} (1979) 115.

\bibitem{Shklovskii+Efros:75}
\textsc{Shklovskii B.~I.} and \textsc{Efros A.~L.}, \emph{J. Phys.
--Condens.
  Matter} \textbf{8}.

\bibitem{KamAn:99}
\textsc{Kamenev A.} and \textsc{Andreev A.}, \emph{Phys. Rev. {\rm
B}}
  \textbf{60} (1999) 2218.

\bibitem{Efetov+Tschersich:02}
\textsc{Efetov K.~B.} and \textsc{Tschersich A.}, \emph{Europhys.
Lett.}
  \textbf{59} (2002) 114; \emph{Phys. Rev.
{\rm B}}
  \textbf{67} (2003) 174205.

\bibitem{Kurland}
\textsc{Kurland I.~L.}, \textsc{Aleiner I.~L.} and
\textsc{Altshuler B.~L.},
  \emph{Phys. Rev. {\rm B}} \textbf{62} (2000) 14886.


\bibitem{Keld}
\textsc{Keldysh L.~V.},  \emph{Zh.\ Eksp.\ Teor.\ Fiz.}
\textbf{47} (1964)
  1515.

\bibitem{RS:86}
\textsc{Rammer J.} and \textsc{Smith H.}, \emph{Rev. Mod. Phys.}
\textbf{58}
  (1986) 323.

\bibitem{JWM:94}
\textsc{{Jauho} A.-P.}, \textsc{{Wingreen} N.~S.} and
\textsc{{Meir} Y.},
  \emph{Phys. Rev. {\rm B}} \textbf{50} (1994) 5528.


  \bibitem{KisGef:06}
\textsc{Kiselev M.~N.} and \textsc{Gefen Y.}, \emph{Phys. Rev.
Lett.} \textbf{96}
  (2006) 066805.


\end{thebibliography}

\end{document}